# Auction-based Efficient Communications in LEO Satellite Systems

Lin Cheng, *Senior Member*, *IEEE*
Next-Gen Systems
CableLabs
Louisville, CO, USA
l.cheng@cablelabs.com

Bernardo A. Huberman, *Fellow*, *AAAS*, *APS*
Next-Gen Systems
CableLabs
Louisville, CO, USA
b.huberman@cablelabs.com

*Abstract*—We propose an auction-based mechanism to improve the efficiency of low earth orbit satellite communication systems. The mechanism allows the ground stations to bid for downlink resources such as spectrum, satellite links, or radios, without the need to send channel status back to satellites. Simulation and experimental results show that this mechanism improves total channel capacity by dynamically leveraging the diversity among satellite-station links; reduces uplink overhead by providing lightweight and effective channel status feedback; simplifies computational complexity and improves scalability; and also provides implicit resource information stemming from the auction dynamics. This new operation mechanism provides a feasible solution for low earth orbit satellites which are sensitive to power consumption and overheating.

*Keywords—auction, low earth orbit satellite, wireless communication*

## I. Introduction

Low earth orbit (LEO) satellite communication networks are starting to pervade people's lives. Although the concept is not new and the cost is still high, developments over the past ten years have made the implementation of LEO satellite networks possible, thanks to a massive reduction in satellite launch costs and new technologies for inter-satellite high bandwidth communications. The satellite Internet is an important component of the future 6G communication protocols [1][2]. In order to provide high-speed and low-latency broadband Internet services, numerous mega constellation plans have been proposed in recent years. The most prominent projects are SpaceX's Starlink [3], Amazon's Project Kuiper [4], and the OneWeb constellation [5], each planning to build networks consisting of hundreds to thousands of LEO satellites. SpaceX plans to build a Starlink constellation composed of about 42000 LEO satellites and to provide communication services with 50 – 150 Mbps data rate and 20 – 40 ms delay [6]. Amazon also plans to build the Kuiper constellation, composed of 1980 LEO satellites and to provide communication services with up to 595 Mbps data rate and 32ms delay. Lastly, China Satellite Network Group, established on April 28, 2021, has integrated domestic mega constellation plans and resources including Hongyun and Hongyan, aimed at building a Chinese version of Starlink [7][8].

As LEO satellites run at altitudes around 500-1000 kilometers, the networks improve the achievable latency and bandwidth as compared with traditional GEO satellite networks, making them viable alternatives to land-based and terrestrial networks. In fact, from a quality of service perspective, satellite networks might ultimately have the advantage over fiber networks as electromagnetic waves propagate faster in vacuum than in fiber optic cables, while the lack of geographic obstructions allows more direct communication paths [9].

The financial feasibility of mega-constellations of LEO satellites relies on reduced launching and manufacturing costs, increasing demand, and improved satellite performance such as digital payloads, steerable multi-beam antennas, advanced modulation and coding schemes, and frequency reuse strategies [10]. Most importantly, LEO satellites are very sensitive to power consumption due to their battery capacity and weight. Power consumption also relates to overheating which is another critical problem due to the nature of low orbits. An efficient link between satellites and ground stations (GS) with high power efficiency, low overhead, and optimized resource allocation among satellite links, frequency spectrum, and antennae is most important to make LEO satellite coverage cost-effective [11].

An optimal or maximal-capacity communication link between satellites and GSs allocates each resource unit to the comparatively strongest GS with constraints such as demands and fairness. An optimal link can effectively save the transmit power from LEO satellites. However, such allocations are usually derived from complicated algorithms and sometimes NP-hard, so that even a satisficing solution requires high computational power and overheating [12][13]. In addition, the optimization of downlink allocations highly relies on satellites' knowledge of GSs' receiving status. Optimal allocation needs complete channel status fed back to satellites, causing a large uplink overhead.

In this paper, we propose an auction-based mechanism for LEO satellite systems. Different from other analytical and optimization approaches, the auction-based mechanism proposed in this paper:

- improves total channel capacity and reduces transmit power by dynamically leveraging the diversity among satellite-station links;
- reduces uplink overhead by providing lightweight and effective channel status feedback;

- simplifies the computational complexity at LEO satellites and improves scalability;
- and also enables the derivation of implicit resource information from auctions.

The method is believed to be a promising strategy for efficient communications between satellites and GSs and make LEO satellite system more cost-effective.

## II. Auction-based Operation

In order to explain our method we will use the allocation of frequency sub-channels (SC) or subcarriers as an example. SCs experience multipath induced frequency-selective fading and frequency-selective interference, leading to different capacities on different SC resources. However, please note that this method may also apply in the management of other components such as radio links, MIMO beams, transmitters and their antennae, or even satellites.

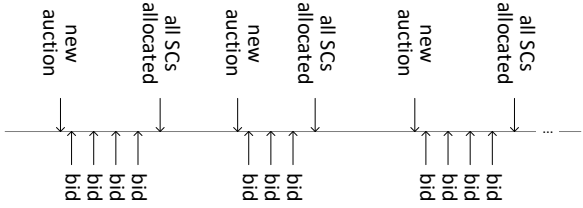

Fig. 1. Timeline of auctions.

Each auction consists of multiple rounds of bidding at the end of which the allocation of SCs is determined. The auctions occur periodically over time and their period length is determined by the nature of channel fading and overhead requirement. For example, fast fading channels require short auction periods; low overhead requires long auction periods. As shown in Fig. 1, a new auction starts at the beginning of each period.

The algorithm is illustrated in Fig. 2. Assume there are $N$ GSs sharing $S$ downlink SCs.

Step 0: At the beginning of an auction, new funds are added to GSs' accounts that are kept at the satellite. These accounts may or may not be accessible to the GSs. The amounts of the funds are determined by other criteria such as demands, priorities, service levels, etc. Please note that fund exchange does not necessarily involve monetary behavior from the GSs or users. In other words, the fund assigned to a GS is solely a variable for the sake of the auction mechanism. The status of all the $S$ SCs are reset to "available" at this step. The GSs also have channel estimates (complex coefficients, SNR, or received power) available for the SCs.

Step 1: Each GS estimates the capacity of each SC based on its channel estimates. The capacity of unavailable SCs is set to zero. Each GS sorts the SCs in the order of the capacities. Denote the SCs as $c_1, c_2, \ldots c_S$ before sorting. After sorting, the SCs are indexed as $c_{n,i_1}, c_{n,i_2}, \ldots c_{n,i_S}$ at GS-$n$, $1 \leq n \leq N$. SC $c_{n,i_1}$ has the highest capacity of all SCs at GS-$n$. GS-$n$ selects $S'_n$ SCs that have the highest capacity. These $S'_n$ SCs are the SCs to bid for the current round of bidding. The number of SCs $S'_n$ is determined by the capacity distribution of SCs. An example of how $S'_n$ is determined is

$$S'_n = \underset{j}{\operatorname{argmax}}\left(\sum_{q=1}^{j} C_{n,i_q} - \frac{j \sum_{q=2}^{S} C_{n,i_q}}{S-1} + \frac{\sum_{q=1}^{S} C_{n,i_q} - S C_{n,i_1}}{S-1}\right) \quad (1)$$

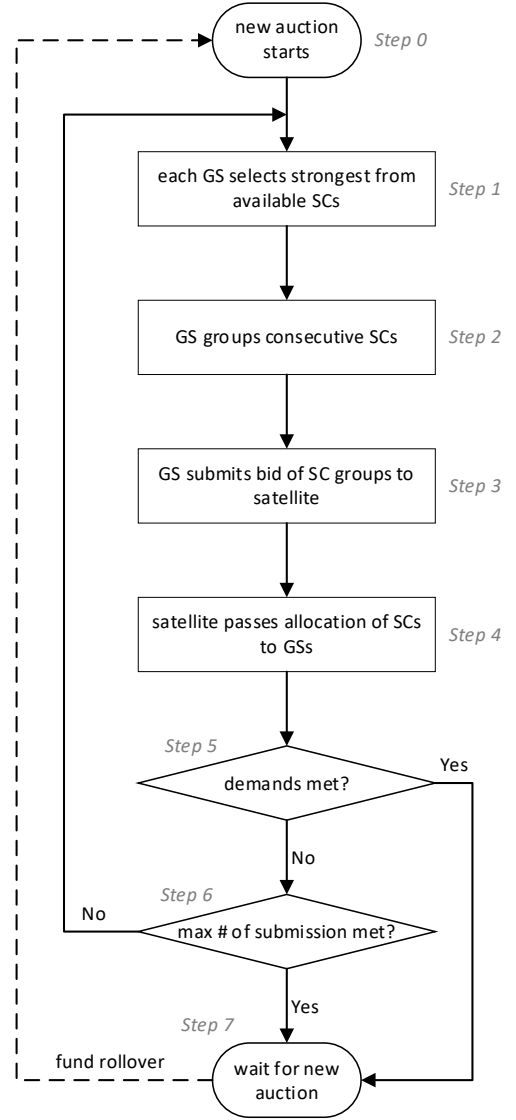

Fig. 2. Steps of auction.

where $C_{n,i_q}$ is the capacity of SC $c_{n,i_q}$, $1 \leq q \leq S$, at GS-$n$. Determining $S'_n$ in such a way can avoid bidding on unavailable SCs. It also allows GSs with strong fading automatically put higher bids on good SCs while GSs with flatter channels automatically back up during first rounds of bidding – both spectral efficiency and fairness play a role in the auction. $S'_n$ is also bound by parameters such as demands, fairness, etc.

Step 2: Thanks to the nature of fading and interference, the selected SCs by each GS in Step 1 are mostly fragmentary, i.e., selected SCs can be divided into a small number of groups and in each group SCs have continuous indices, as shown in Fig. 3. Assume GS-$n$ has $G_n$ groups of selected subcarriers, $\mathbf{G}_{n,1}, \mathbf{G}_{n,2}, \ldots \mathbf{G}_{n,G_n}$. In the example in Fig. 3, $G = 2$ for all three GSs, i.e., every GS bid for two groups of SCs. The GS assigns bid prices to these groups. An example of bid price assignment is as follows. Each GS-$n$ first assigns a ratio

$$b_{n,g} = \sum_{i \in \mathbf{G}_{n,g}} C_{n,i} / \sum_{i \in \cup_g \mathbf{G}_{n,g}} C_{n,i} \qquad (1)$$

to each group $\mathbf{G}_{n,g}$. GS-$n$ sends this ratio, $b_{n,g}$, to the satellite. The satellite counts the bid price of each of the SCs in $\mathbf{G}_{n,g}$ from GS-$n$ as

$$\frac{b_{n,g}}{|\mathbf{G}_{n,g}|} M_n \qquad (1)$$

where $M_n$ is the current total amount of fund in GS-$n$'s account, $|\cdot|$ denotes the total SC number in a group.

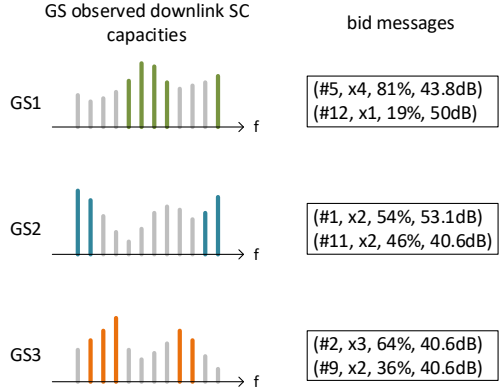

Fig. 3. An example to illustrate GSs sending bid messages bidding for groups of SCs.

Step 3: The GSs send their bids to the satellite through the control channel in uplink. An example of the bid messages is shown in Fig. 3. Each group of SCs to bid contains four mandatory fields: index of starting SC, total SC number in the group $|\mathbf{G}_{n,g}|$, ratio $b_{n,g}$, and the lowest capacity (or SNR, or spectral efficiency, or power) of the SCs in the group.

Step 4: The satellite allocates each SC that receives at least one bid to the GS that offers the highest bid. The allocation should meet other constraints such as maximal number of SCs per GS allowed, airtime fairness, etc. The satellite also updates that GS's account by deducting a certain amount from the account. An example of how much is charged to the account is the bid price, i.e., $\frac{b_{n,g}}{|\mathbf{G}_{n,g}|} M_n$. Another example is the second highest bid price [14], if the second highest bid price exists. The satellite then sends the allocation (and optionally the account balance) information of successfully sold SCs to GSs through the control channel in downlink. GSs set the status of all the sold SCs to "unavailable".

Step 5: The satellite decides on whether each GS's capacity demand is met based on the downlink traffic that has been or to be sent.

Step 6: Each system decides if the total number of rounds of bidding has exceeded a certain number.

Step 7: The system continues or carries out data transmission using the final SC allocation for downlink. GSs' account may or may not reset to zero, i.e., rollover or not.

It is worth noting that downlink payload data transmission does not have to wait until the end of the auction process to start. The downlink transmission proceeds with the allocation from the previous auction result and then updates its allocation once the ongoing auction finishes. In other words, auctions proceed and update the allocation with the control channel and data channel going on concurrently.

III. SIMULATION AND EXPERIMENT

An example of Python simulation of the method is demonstrated in this section. Fig. 4 shows an example of how SCs are allocated at the end of an auction, i.e., Step 7, from the simulation. 16 GSs and 1024 SCs are included. Downlink transmission experiences frequency selective fading caused by multipath and interference.

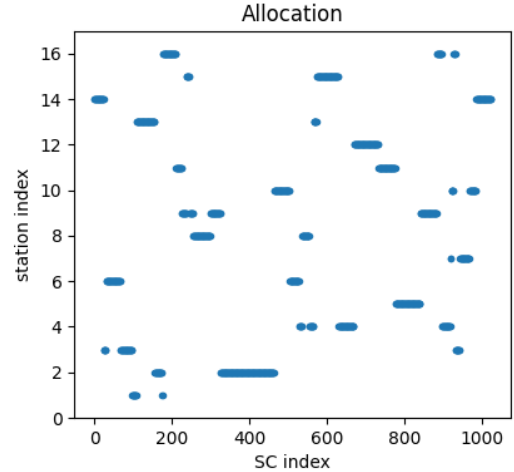

Fig. 4. Simulation result: SC allocation at the end of an auction for 16 GSs and 1024 SCs. (240-MHz total bandwidth, regional interference at #300 - #400 SC, maximum 4 rounds of bidding per auction)

A. Efficiency Improvement

The method improves communication efficiency compared with random allocations, which do not take resource diversity into consideration. The blue points in Fig. 5 show an example of the average capacity of the 16 GSs provided by 100 auction trials, i.e., run Steps of Auction for 100 times for 100 instances of channel models, from the simulation. The orange points show the corresponding result of allocation where SCs are allocated to GSs randomly. On average the method shows a 29% improvement in this case, subject to other parameters such as channel fading, interference strength, GS diversity, etc. In addition, the green points show the capacity limit of the system, i.e., a satellite assigns each SC to the absolute strongest GS without any fairness. Our auction-based method has a small decrease relatively, while it guarantees much better fairness which can be examined by measuring the standard deviation of the capacity among the 16 GSs, as shown in Fig. 6.

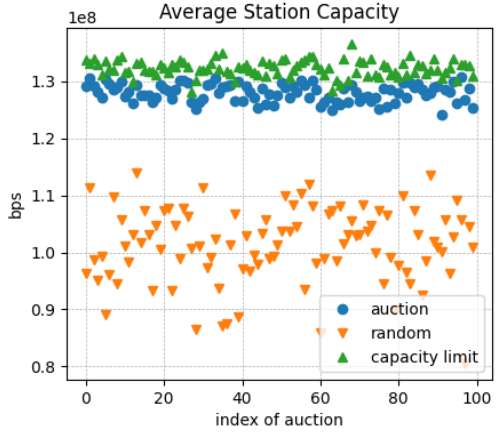

Fig. 5. Simulation result: average capacity of 16 GSs from 100 auctions, with comparison with two references.

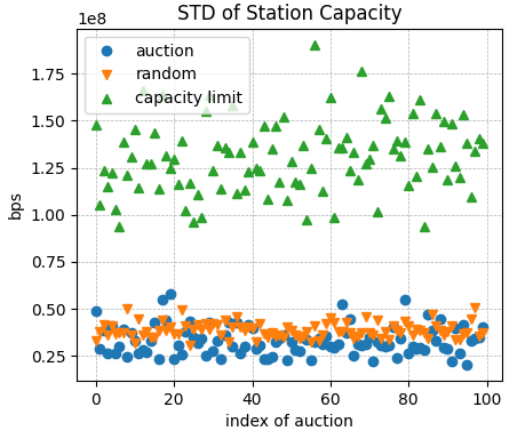

Fig. 6. Simulation result: standard deviation of capacity of 16 GSs from 100 auctions, with comparison with two references.

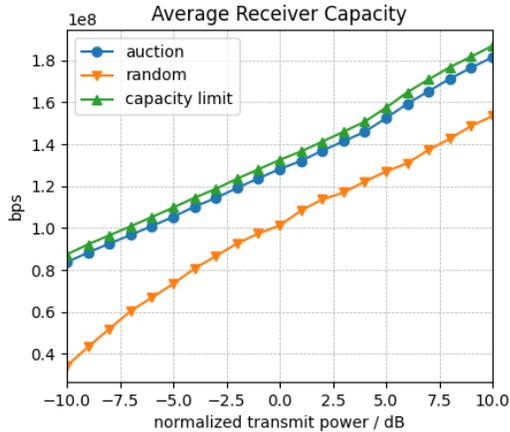

Fig. 7. Simulation result: average capacity of 16 GSs versus normalized satellite transmitted downlink power over 100 auctions, with comparison with two references.

We also tested the method's performance on average capacity versus normalized transmitted downlink power, as shown in Fig. 7, in order to examine the fidelity of the method under different SNR conditions. Consistent with Fig. 5, the method has obvious improvement compared with randomly allocating SCs to GSs while having a small gap from capacity limit. On average, there is a 5-dB difference between the proposed method and randomly assigning SCs. This is promising power saving for LEO satellites.

### B. Overhead Reduction

Theoretically, this methodology has a downlink overhead similar to other methods. On the other hand, our mechanism reduces uplink overhead compared with systems that feedback complete channel status. To feedback complete channel status, each GS needs to pass a large amount of information containing complex coefficients, SNR, or received power of each SC to the satellite. For 1024 subcarriers as an example, the same as what we use in the simulation, assuming the channel status for each SC takes 10 information bits to describe the channel characteristic, the total uplink traffic is ~10K information bits per user every time the SC allocation updates.

In our method, assuming each bid message takes 40 bits (10 bits each field) in Step 3, Fig. 8 shows the uplink traffic in bit/auction/GS to show the overhead occupancy of the uplink control channel provided by 100 auction trials from the simulation. The average overhead for each GS every update is 214 bit. Compared with ~10Kbit per GS per update, the method provides significant overhead reduction in uplink while performing similarly to a complete-information channel status feedback.

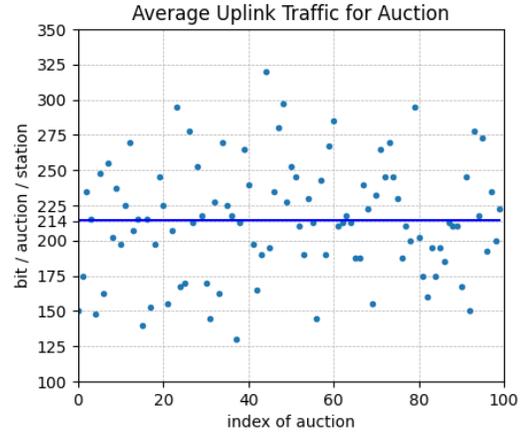

Fig. 8. Simulation result: uplink overhead from 100 auctions, average at 214 bit/auction/GS.

### C. Low Computational Complexity

The method simplifies computation complexity compared with optimal allocation. Optimal allocation, as aforementioned, is NP-hard while the computational complexity of this method is $O(S)$ at the satellite and $O(S \log^2 S)$ at each GS. This also makes the method scale more feasibly when the number of GSs and/or the number of SCs increase. Most importantly, the majority part of the computation is processed at the GSs instead of at the satellites, causing less power consumption and overheating at the satellites.

## D. Experiment Setup

To further verify our idea, we built a software-defined radio (SDR) setup with three transmitters (behaving as satellite transmitters) and two receivers (behaving as two GSs), as shown in Fig. 9. The setup also includes two interference sources that bring higher diversity. The transmitters and receivers run the auction mechanism in GNU Radio. We tested its performance on average capacity versus normalized transmitted downlink power over 100 auctions. As shown in Fig. 10, the method has obvious improvement compared with randomly allocating SCs to receivers while has a small gap from capacity limit. This experiment result is consistent with the simulation result in Fig. 7. A 4.7-dB power saving is observed on average. Although it is still a promising saving, it is lower than the 5-dB saving in the simulation result especially at a high-power region due to the lower number of receivers in the experiment setup.

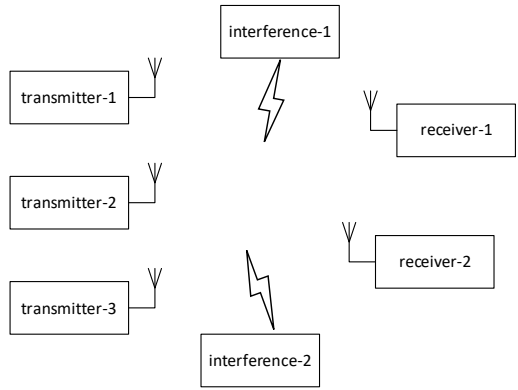

Fig. 9. SDR experiment setup. (512 OFDM SCs, 30-MHz total bandwidth)

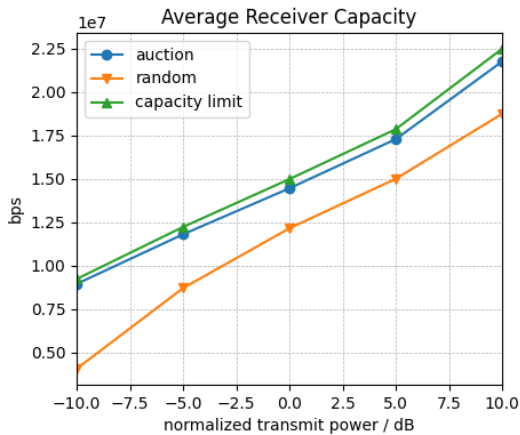

Fig. 10. Experiment result: average capacity of 2 receivers versus normalized transmitted downlink power over 100 auctions, with comparison with two references.

## IV. CHANNEL STATUS IMPLICATION

The method enables the derivation of implicit channel information from the auctions. There are multiple scenarios and applications of this feature of the method. Two examples are as follows.

## A. Unsold SCs

Depending on the setup of the realization of the method, there may be a small number of unsold SCs at the end of an auction. Instead of leaving these SCs idle, the satellite can allocate these SCs to GSs for free. The question is which GS to allocate to so that the SCs can have their best utility in terms of spectral efficiency. To answer this question, we consider the example in Fig. 3 where we assume the first and last SCs are unsold. The satellite checks the record of the GSs' bid messages and found out that GS-2 bid for the second SC. Considering the nature of channel fading, the capacity of the first SC at GS-2 should be similar to that of the second SC which GS-2 bid as one of its strongest SCs. Therefore, it makes sense to allocate the first SC to GS-2. Both GS-2 and GS-1 have records of bidding for the second to last SC. Again, considering the nature of channel fading, the last SC should be relatively strong compared with other unwanted SCs for these two GSs. After further comparing the SNR field of the bid messages from these two GSs, the satellite decides to allocate this SC to GS-1 as it has a 50-dB SC group next to the unsold SC.

## B. Interference Detection

If a SC is left unsold repeatedly over many auctions, it is a sign that there may exist regional interference over these SCs. In this case, the satellite may decide to keep these SCs idle and adjust its power loading, i.e., power is unloaded from these interfered SCs to other useful SCs.

Other than monitoring the sale record, a better way to detect regional interference is to monitor the transaction price over auctions. To show how it works, we added interference by decreasing the SNR of SC #300 – #400 among the 1024 SCs for all the 16 GSs in the simulation. After 100 auction trials, Fig. 11 shows the transacted prices averaged over the 100 auctions. These prices clearly imply which SCs are interfered with. Further power loading can be implemented based on this implication to improve power efficiency.

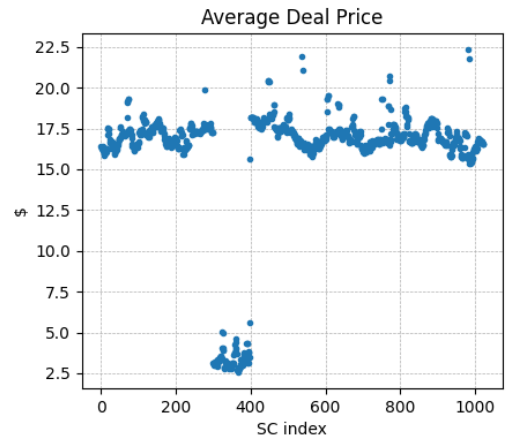

Fig. 11. Simulation result: average deal price of 1024 SCs revealing the frequency of interference.

We point out that monitoring interference using deal price provides higher accuracy than monitoring using bid messages. This is because it is very likely that an interfered SC is still bidden for and sold as a "cheap merchandise" at the end of an auction after no GSs bid for it for the first several rounds of bidding. As shown, for example, in Fig. 4, the interfered SC #301 – #400 are successfully sold to GS-2, 8, and 9 although at much lower prices.

## V. Discussion

The majority of our paper uses point-to-multipoint communication as an example whereas the method also applies in multipoint-to-multipoint communication. In fact, in our experiment aforementioned, we applied the method in a multipoint-to-multipoint setup. We have used the allocation of SCs as the example whereas the proposed operation mechanism also applies in the management of spatial streams, radios, antennae, satellites, etc.

## VI. Conclusion

In this paper, we have proposed an auction-based mechanism to improve the efficiency of low earth orbit satellite communication systems. The auction process allows the ground stations to bid for downlink resources such as spectrum, satellite links, or radios, without the need of sending redundant channel status back to satellites and thus save uplink overheads. Simulations and experimental results demonstrate that the proposed mechanism improves total channel capacity and reduces transmit power by dynamically leveraging the diversity among satellite-station links; reduces uplink overhead by providing lightweight and effective channel status feedback; simplifies the computational complexity at LEO satellites and improves scalability; and also enables the derivation of implicit resource information from auctions. This operation mechanism provides a feasible solution for low earth orbit satellites that are sensitive to power consumption and overheads.

References

[1] B. Deng, C. Jiang, H. Yao, S. Guo and S. Zhao, "The Next Generation Heterogeneous Satellite Communication Networks: Integration of Resource Management and Deep Reinforcement Learning", in *IEEE Wireless Communications*, vol. 27, no. 2, pp. 105-111, Apr. 2020.

[2] X. Zhu, C. Jiang, L. Kuang, Z. Zhao, and S. Guo, "Two-Layer Game Based Resource Allocation in Cloud Based Integrated Terrestrial-Satellite Networks", in *IEEE Transactions on Cognitive Communications and Networking*, vol. 6, no. 2, pp. 509-522, Jun. 2020.

[3] Starlink. (2019) Starlink. [Online]. Available: https://www.starlink.com/

[4] Amazon. (2020) Project Kuiper. [Online]. Available: https://www.amazon.jobs/en/teams/projectkuiper

[5] OneWeb. (2019) Oneweb. [Online]. Available: https://www.oneweb.world/

[6] G. Zeng, Y. Zhan and X. Pan, "Failure-Tolerant and Low-Latency Telecommand in Mega-Constellations: The Redundant Multi-Path Routing," in *IEEE Access*, vol. 9, pp. 34975-34985, 2021.

[7] H. Xie, Y. Zhan, G. Zeng and X. Pan, "LEO Mega-Constellations for 6G Global Coverage: Challenges and Opportunities," in *IEEE Access*, vol. 9, pp. 164223-164244, 2021.

[8] G. Zeng, Y. Zhan, H. Xie, C. Jiang, "Networked Satellite Telemetry Resource Allocation for Mega Constellations," in *IEEE International Conference on Communications*, ICC, 2022.

[9] B. Kempton and A. Riedl, "Network Simulator for Large Low Earth Orbit Satellite Networks," in *IEEE International Conference on Communications*, ICC, 2021.

[10] N. Pachler, I. del Portillo, E. F. Crawley, B. G. Cameron, "An Updated Comparison of Four Low Earth Orbit Satellite Constellation Systems to Provide Global Broadband," in *IEEE International Conference on Communications*, 2021.

[11] I. Bisio and M. Marchese, "Power Saving Bandwidth Allocation over GEO Satellite Networks", in *IEEE Communications Letters*, vol. 16, no. 5, pp. 596-599, May 2012.

[12] P. Huang, Y. Gai, B. Krishnamachari and A. Sridharan, "Subcarrier Allocation in Multiuser OFDM Systems: Complexity and Approximability," *IEEE Wireless Communication and Networking Conference*, Sydney, NSW, 2010, pp. 1-6, 2010.

[13] H. A. Ammar, R. Adve, S. Shahbazpanahi, G. Boudreau, and K. V. Srinivas, "Downlink Resource Allocation in Multiuser Cell-Free MIMO Networks With User-Centric Clustering", in *IEEE Transactions on Wireless Communications*, vol. 21, No. 3, pp. 1482-1497, Mar. 2022.

[14] I. Caragiannis, C. Kaklamanis, P. Kanellopoulos, M. Kyropoulou, B. Lucier, R. Paes Leme, E. Tardos, "Bounding the inefficiency of outcomes in generalized second price auctions," *Journal of Economic Theory*, 156: 343–388, 2015.